\documentclass[conference]{IEEEtran}
\IEEEoverridecommandlockouts
\usepackage{cite}
\usepackage{amsmath,amssymb,amsfonts}
\usepackage{algorithmic}
\usepackage{graphicx}
\usepackage{textcomp}
\usepackage{xcolor}
\usepackage{tikz}
\def\BibTeX{{\rm B\kern-.05em{\sc i\kern-.025em b}\kern-.08em
    T\kern-.1667em\lower.7ex\hbox{E}\kern-.125emX}}
\usetikzlibrary{3d,decorations.text,shapes.arrows,positioning,fit,backgrounds, calc}

\tikzset{pics/fake box/.style args={% #1=color, #2=x dimension, #3=y dimension, #4=z dimension
#1 with dimensions #2 and #3 and #4}{
code={
\draw[gray,ultra thin,fill=#1]  (0,0,0) coordinate(-front-bottom-left) to
++ (0,#3,0) coordinate(-front-top-right) --++
(#2,0,0) coordinate(-front-top-right) --++ (0,-#3,0) 
coordinate(-front-bottom-right) -- cycle;
\draw[gray,ultra thin,fill=#1] (0,#3,0)  --++ 
 (0,0,#4) coordinate(-back-top-left) --++ (#2,0,0) 
 coordinate(-back-top-right) --++ (0,0,-#4)  -- cycle;
\draw[gray,ultra thin,fill=#1!80!black] (#2,0,0) --++ (0,0,#4) coordinate(-back-bottom-right)
--++ (0,#3,0) --++ (0,0,-#4) -- cycle;
\path[gray,decorate,decoration={text effects along path,text={CONV}}] (#2/2,{2+(#3-2)/2},0) -- (#2/2,0,0);
}
}}
\tikzset{circle dotted/.style={dash pattern=on .05mm off 2mm,
                                         line cap=round}}
\tikzstyle{arrow} = [thick,->,>=stealth, line width=1.5pt]
\begin{document}

\title{Detection of AI Synthesized Hindi Speech}

\author{\IEEEauthorblockN{Karan Bhatia  \IEEEauthorrefmark{1},
Ansh Agrawal \IEEEauthorrefmark{1},
Priyanka Singh\IEEEauthorrefmark{1}, and
Arun Kumar Singh\IEEEauthorrefmark{2}}
\IEEEauthorblockA{\IEEEauthorrefmark{1}Dhirubhai Ambani Institute of Information and Communication Technology\\
Gandhinagar, Gujarat, India\\ 
Email: \{201801417,201801110,Priyanka\_Singh\}@daiict.ac.in} 
\IEEEauthorblockA{\IEEEauthorrefmark{2} Indian Institute of Technology Jammu,\\
Jammu, India\\
Email: singh.arun.kumar@ieee.org}}

% \author{\IEEEauthorblockN{Priyanka Singh\IEEEauthorrefmark{1},
% Abhishek Singh\IEEEauthorrefmark{2},
% Gabriel Cojocaru\IEEEauthorrefmark{2}, and
% Ramesh Raskar\IEEEauthorrefmark{2}}
% \IEEEauthorblockA{\IEEEauthorrefmark{1}Dhirubhai Ambani Institute of Information and Communication Technology\\
% Gandhinagar, Gujarat, India\\ Email: Priyanka\_Singh@daiict.ac.in}
% \IEEEauthorblockA{\IEEEauthorrefmark{2}MIT Media Lab\\
% Cambridge, MA, USA\\
% Email: \{abhi24,r\_eality,raskar\}@mit.edu}}
% \author{\IEEEauthorblockN{1\textsuperscript{st} Given Name Surname}
% \IEEEauthorblockA{\textit{dept. name of organization (of Aff.)} \\https://www.overleaf.com/project/6123a8527ab31d89d71c7fe2
% \textit{name of organization (of Aff.)}\\
% City, Country \\
% email address or ORCID}
% \and
% \IEEEauthorblockN{2\textsuperscript{nd} Given Name Surname}
% \IEEEauthorblockA{\textit{dept. name of organization (of Aff.)} \\
% \textit{name of organization (of Aff.)}\\
% City, Country \\
% email address or ORCID}
% \and
% \IEEEauthorblockN{3\textsuperscript{rd} Given Name Surname}
% \IEEEauthorblockA{\textit{dept. name of organization (of Aff.)} \\
% \textit{name of organization (of Aff.)}\\
% City, Country \\
% email address or ORCID}
% \and
% \IEEEauthorblockN{4\textsuperscript{th} Given Name Surname}
% \IEEEauthorblockA{\textit{dept. name of organization (of Aff.)} \\
% \textit{name of organization (of Aff.)}\\
% City, Country \\
% email address or ORCID}
% }

\maketitle

\begin{abstract}
The recent advancements in generative artificial speech models have made possible the generation of highly realistic speech signals. At first, it seems exciting to obtain these artificially synthesized signals such as speech clones or deep fakes but if left unchecked, it may lead us to digital dystopia. One of the primary focus in audio forensics is validating the authenticity of a speech. Though some solutions are proposed for English speeches but the detection of synthetic Hindi speeches have not gained much attention. Here, we propose an approach for discrimination of AI synthesized Hindi speech from an  actual human speech. We have exploited the Bicoherence Phase, Bicoherence Magnitude, Mel Frequency Cepstral Coefficient (MFCC), Delta Cepstral, and Delta Square Cepstral as the discriminating features for machine learning models. Also, we extend the study to using deep neural networks for extensive experiments, specifically VGG16 and homemade CNN as the architecture models. We obtained an accuracy of 99.83\% with VGG16 and 99.99\% with homemade CNN models.

%With the recent advancements in generative artificial speech models, generation of highly realistic speech signals have been possible. This also raises alarming concerns that can propagate as speech clones or deep fakes. One of the primary problems in digital forensics is validating the authenticity of a speech. There has been no recent significant contribution for the detection of synthetic Hindi language and simultaneously there has been a rise of synthetic speech generators for Hindi speech. We propose an approach to tackle this problem by training AI models on multiple speech features. 
\end{abstract}


\section{Introduction}
The field of speech forensics has progressed a lot but very few schemes have been proposed for detection of AI synthesized speech. Some techniques address speech spoofing and tampering \cite{jelil2017spoof} based on instantaneous frequency and cepstral features,  but they are not
explicit for detecting AI synthesized speech. A comparison of features for synthetic speech detection is presented in \cite{hanilcci2015classifiers}. While synthesizing AI speech, first-order Fourier coefficients or second-order power spectrum correlations can be easily tuned to match a human speech but it's comparatively much harder for third-order bi-spectrum correlations \cite{farid19}. Thus, higher order correlations are used to 
discriminate between human and AI speech. Mel spectral analysis revealed the fact that a durable power component is missing in the AI synthesized speeches  which is present in the human speech \cite{arun21}. This durable power component is attributed to the vocal tract present in humans \cite{muda2010voice}. Also, 
$\Delta-Cepstral$ and $\Delta^{2}-Cepstral$ are two other discriminatory features related to Mel Frequency Cepstral Coefficient (MFCC) \cite{kumar2011deltacepstral}.
We have exploited these established features, specifically Bicoherence Phase, Bicoherence Magnitude, MFCC, Delta Cepstral, and Delta Square Cepstral as the discriminating features for human versus AI synthesized  speeches for our machine learning models.  There are some works related to this issue, done primarily for the English and Mandarin languages but Hindi has not gained much attention \cite{wu2013temporal} \cite{farid19}. In our work, we  primarily focus on the Hindi speeches. A major bottleneck  while working on this problem was the lack of standard datasets for Hindi speech. We contributed our own dataset to carry out this study. For this, we considered various freely available text to speech converters that produced natural sounding audio signals and collected synthetic speech samples from them. \\

The rest of the paper is organized as follows: the detailed description of the dataset is provided in section \ref{sec:dataset} followed by the experimentation details of the machine learning methods in section \ref{sec:ML} and deep learning architectures in section \ref{sec:DL}. The final results are discussed in section \ref{sec:results} followed by conclusion in section \ref{sec:conclusion}.

\section{Data set} \label{sec:dataset}
For our work, we collected human and synthetic speech samples from various sources and constructed a data set. We collected a dataset of total 12,890 samples of 5s each in length. For human speech, we had a total of 8,140 samples, out of which we collected 5,750 audio samples from a existing dataset compiled by Indian Institute of Technology, Madras \cite{iit20} and the remaining 2,390 audio samples were recorded by microphone. For AI synthesised speech we collected 4,750 audio  samples from four different text-to-speech synthesizers: IITM TTS, Hearling, Amazon Polly and Voice Maker. These TTS synthesizers are freely available and thus can be used by anyone. We chose these TTS synthesizers since audio generation was simple and quick and also the audio generated was natural sounding. To maintain diversity in the dataset we collected both male and female voices for human and AI synthesised speech.

%Initially, we had created a data set of 1,100 samples where 880 samples were of AI synthesised speech (220 for each class) and the remaining 220 samples were for human speech.

\section{Experiments using Machine Learning Models}\label{sec:ML}
Prior to using the Machine Learning (ML) models, we compute the discriminatory features. First, we calculate the bicoherence magnitude and phase for all the speech samples. Then we calculate the mean, variance, skewness, and kurtosis for both magnitude and phase which is given by:
\begin{itemize}
    \item Mean , $\mu_{X} = E_{X}[X]$ \\
    \item Variance , $\sigma_{X} = E_{X}[(X-\mu_{X})^{2}]$\\
    \item Skewness , $\gamma_{X} = E_{X}[(\frac{X-\mu_{X}}{\sigma_{X}})^{3}]$\\
    \item Kurtosis , $\kappa_{X} = E_{X}[(\frac{X-\mu_{X}}{\sigma_{X}})^{4}]$\\
\end{itemize}
where $E_{X}[.]$ is the expected value operator and X is the random variable. For magnitude, we consider X = M and for phase, X = P. We then calculate the four statistical moments by replacing the expected value operator with the average.
We also calculate the mean and variance for MFCC, $\Delta-Cepstral$ and $\Delta^{2}-Cepstral$. We get a 15-D feature vector where the first 8 entries represent the above four statistical moments for magnitude and phase. The next 6 entries represent the mean and variance of MFCC, $\Delta-Cepstral$ and $\Delta^{2}-Cepstral$. The last entry represents the class of the audio sample. This differs based on the classification. If we consider binary-class then the last entry consists of 2 types: Human or AI synthesised speech. Binary classification is also the main focus of the project. If we consider multi-class, then the last entry consists of 5 types: Human, IITM TTS, Hearling, Amazon Polly and Voice Maker. 
We first tested the accuracy individually for Bicoherence magnitude, Bicoherence Phase, MFCC, $\Delta-Cepstral$ and $\Delta^{2}-Cepstral$. Then, we combined Bicoherence Magnitude and Phase and finally we tested by combining all the features together. This helped us visualize the impact on the accuracy of the different features. For both binary-class and multi-class classification,  we experimented with the following machine learning algorithms: Linear Discriminant, Linear SVM, Weighted KNN, Boosted Trees, Bagged Trees and RUSBoosted Trees. These algorithms were used for training, validation and testing. Validations was done using 5-fold cross-validation.

\begin{table*}[ht]
\centering
\vspace{2mm}
\caption{Binary-Class : Accuracy of individual features and combined features for different machine learning models.}
\begin{tabular}{|c|c|c|c|c|c|c|c|}
\hline
\multicolumn{1}{|l|}{}             & \multicolumn{5}{c|}{\textbf{Individual Features}}                           & \multicolumn{2}{c|}{\textbf{Combined Features}} \\ \hline
\textbf{Various Models} &
  {\begin{tabular}[c]{@{}c@{}}Bicoherence \\ Magnitude\end{tabular}} &
  {\begin{tabular}[c]{@{}c@{}}Bicoherence \\ Phase\end{tabular}} &
  {MFCC} &
  {\begin{tabular}[c]{@{}c@{}}Delta \\ Cepstral\end{tabular}} &
  {\begin{tabular}[c]{@{}c@{}}Delta \\ Square Cepstral\end{tabular}} &
  {\begin{tabular}[c]{@{}c@{}}Bicoherence\\ (Magnitude \& Phase)\end{tabular}} &
  {\begin{tabular}[c]{@{}c@{}}Bicoherence\\(Magnitude \& Phase)\\ \&\\ MFCC\\ \&\\ Delta Cepstral\\ \&\\ Delta Square\\ Cepstral\end{tabular}} \\  
\hline
{Linear Discriminant}      &57.5         & 57.5          &80.9          & 77.4          & 78.5        & 57.4        & 81.3      \\  
\hline
{Linear SVM}        & 57.5          & 57.5          & 18.3      & 82.8   
& 80.7        & 57.5          & 82.6                    \\  
\hline
{Weighted KNN}              & 61.5          & 55.9          & 93.0      & 95.1  
& 90.7        & 59.9                   & 81.9  \\ 
\hline
{Boosted Trees Ensemble}    & \textbf{63.0}       &\textbf{57.5}        
&\textbf{94.0}  &96.4  &\textbf{91.9}        & 62.9             & 98.7      \\ 
\hline
{Bagged Trees Ensemble}     & 63.5          & 55.8          & 93.9 
&95.4  &90.6        &\textbf{64.9}    &\textbf{98.9}      \\ 
\hline
{RUSBoosted Trees Ensemble} & 62.0   & 51.6          & 93.6      &\textbf{96.8} 
& 92.1        & 62.3                   & 98.1                   \\ 
\hline
\end{tabular}
\label{tab:BinaryClass Experiment1}
\end{table*}

\begin{table*}[ht]
\centering
\vspace{2mm}
\caption{Multi-Class : Accuracy of individual features and combined features for different machine learning models.}
\begin{tabular}{|c|c|c|c|c|c|c|c|}
\hline
\multicolumn{1}{|l|}{}             & \multicolumn{5}{c|}{\textbf{Individual Features}}                           & \multicolumn{2}{c|}{\textbf{Combined Features}} \\ \hline
\textbf{Various Models} &
  {\begin{tabular}[c]{@{}c@{}}Bicoherence \\ Magnitude\end{tabular}} &
  {\begin{tabular}[c]{@{}c@{}}Bicoherence \\ Phase\end{tabular}} &
  {MFCC} &
  {\begin{tabular}[c]{@{}c@{}}Delta \\ Cepstral\end{tabular}} &
  {\begin{tabular}[c]{@{}c@{}}Delta \\ Square Cepstral\end{tabular}} &
  {\begin{tabular}[c]{@{}c@{}}Bicoherence\\ (Magnitude \& Phase)\end{tabular}} &
  {\begin{tabular}[c]{@{}c@{}}Bicoherence\\(Magnitude \& Phase)\\ \&\\ MFCC\\ \&\\ Delta Cepstral\\ \&\\ Delta Square\\ Cepstral\end{tabular}} \\  
\hline
{Linear Discriminant}      & \textbf{57.5}         & \textbf{57.5}          &67.2          & 66.4          & 63.2        & 57.5        & 73.8      \\  
\hline
{Linear SVM}                & 57.5          & 57.5          & 28.6      & 21.0   
& 63.5        & 57.5          & 73.1                     \\  
\hline
{Weighted KNN}              & 53.1          & 54.0          & 85.4      & 86.7  
& 78.3        & 55.0                   & 73.8                   \\ 
\hline
{Boosted Trees Ensemble}    & 57.0       & 57.5         & 80.4  &82.4 
&78.2        & 57.0                   & 91.0      \\ 
\hline
{Bagged Trees Ensemble}     & 54.8          & 55.5          & \textbf{86.8} 
&\textbf{86.9}  &\textbf{79.5}        &\textbf{57.7}    &\textbf{ 93.4}      \\ 
\hline
{RUSBoosted Trees Ensemble} & 25.2          & 20.1          & 76.3      &83.6          & 78.5        & 24.7                   & 87.6                   \\ 
\hline
\end{tabular}
\label{tab:MultiClass Experiment1}
\end{table*}

\begin{figure} [ht]
    \centering
    \includegraphics[width=0.45\textwidth]{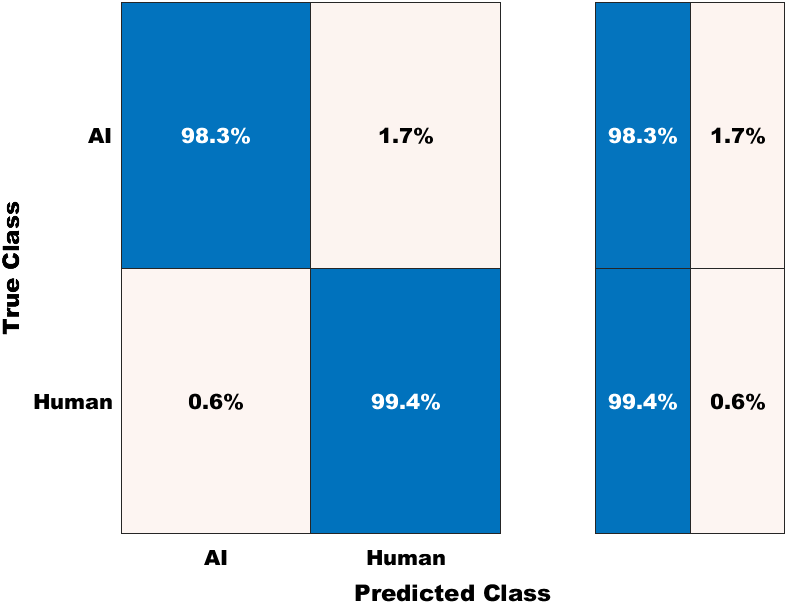}
    \caption{Confusion Matrix: Bagged Trees Model for Binary-Class}
    \label{fig:binaryclasstest}
\end{figure}

\begin{figure} [ht]
    \centering
    \includegraphics[width=0.45\textwidth]{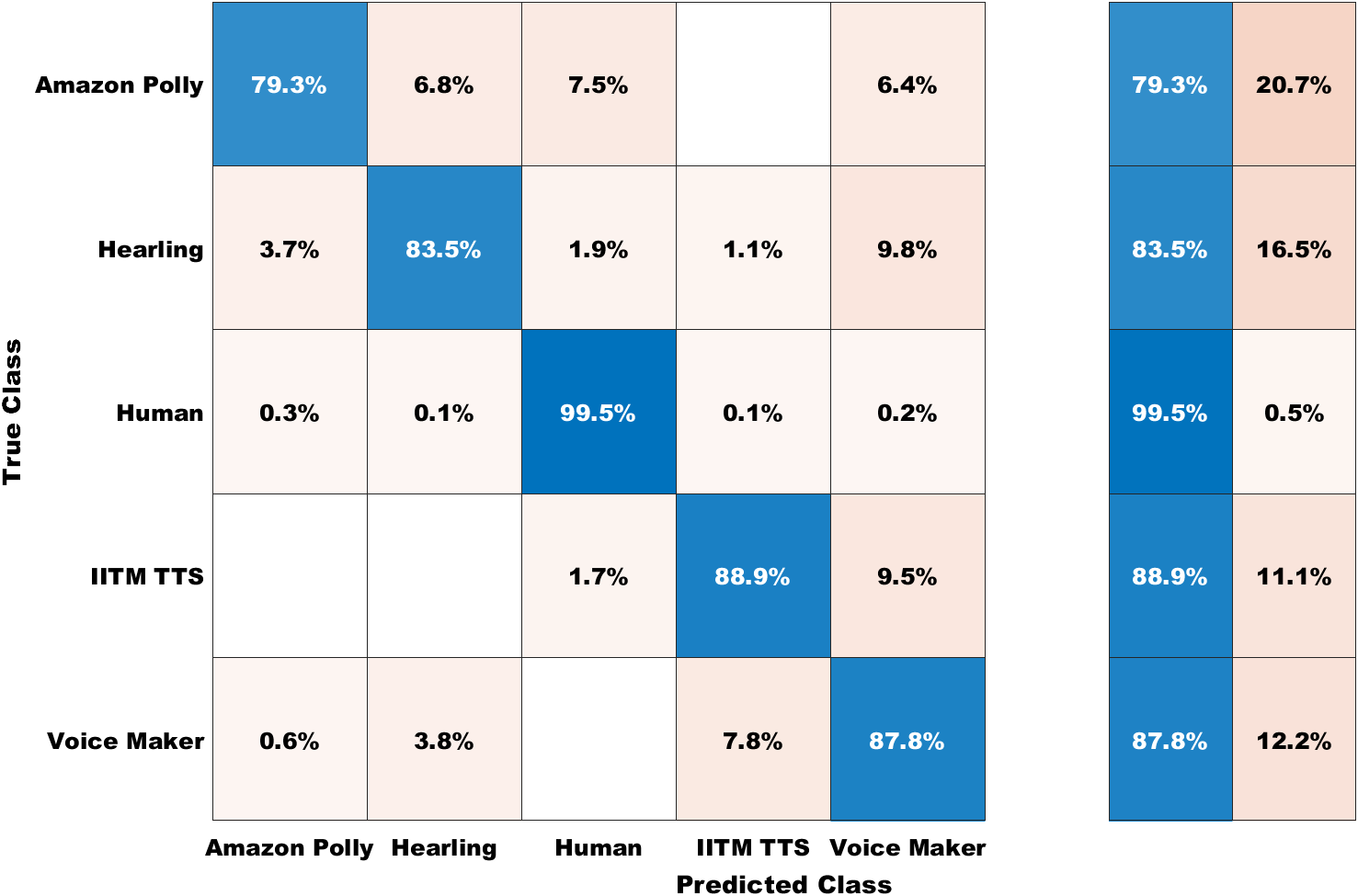}
    \caption{Confusion Matrix : Bagged Trees Model for Multi-Class}
    \label{fig:multiclasstest}
\end{figure}

\section{Experiments using Deep Learning Architectures} \label{sec:DL}
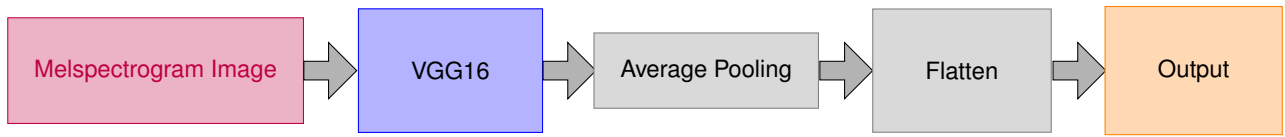
\begin{figure*}[]
	\centering
\begin{tikzpicture}[x={(1,0)},y={(0,1)},z={({cos(60)},{sin(60)})},
font=\sffamily\small,scale=2]
\node[rectangle, draw, purple, fill=purple!30, inner sep = 1em, minimum size =4em] at (4.04, 0.5, 0) {Melspectrogram Image};
\node[draw,single arrow, black, fill=black!30, minimum size =2em] at (5.17, 0.5, 0){};
\node[rectangle, draw, blue, fill= blue!30, inner sep = 2em, text = black] at (6, 0.5, 0) {VGG16};
\node[draw,single arrow, black, fill=black!30, minimum size =2em] at (6.76, 0.5, 0){};
\node[rectangle, draw, gray, fill = gray!30, inner sep = 1em, text = black] at (7.7, 0.5, 0) {Average Pooling};
\node[draw,single arrow, black, fill=black!30, minimum size =2em] at (8.6, 0.5, 0){};
\node[rectangle, draw, gray, fill = gray!30, inner sep = 2em, text = black] at (9.4, 0.5, 0) {Flatten};
\node[draw,single arrow, black, fill=black!30, minimum size =2em] at (10.15, 0.5, 0){};
\node[rectangle, draw, orange, fill = orange!30, inner sep = 2em, text = black] at (10.94, 0.5, 0) {Output};
\end{tikzpicture}
	\caption{Model architecture using pre-trained VGG16. Output has 5 classes}
	\label{fig: VGG16}
\end{figure*}

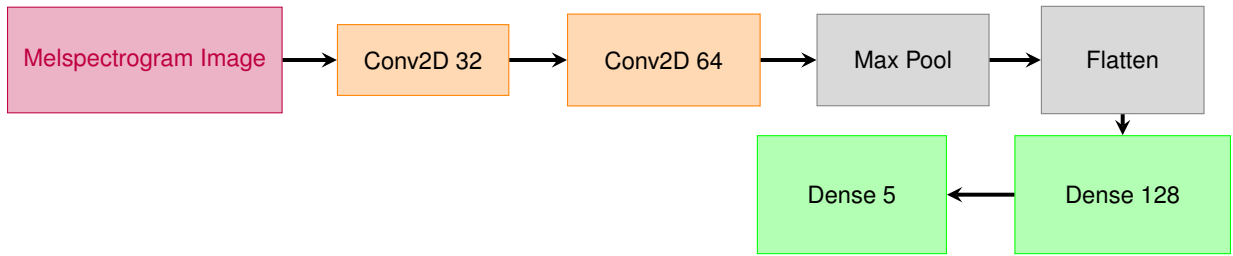
\begin{figure*}[ht]
	\centering
\begin{tikzpicture}[x={(1,0)},y={(0,0)},z={({cos(60)},{sin(60)})},
font=\sffamily\small,scale=2]
\node[rectangle, draw, purple, fill=purple!30, inner sep = 0.6em, minimum size =4em] at (0, 0.5, 0) (melspec) {Melspectrogram Image};
\node[rectangle, draw, orange, fill= orange!30, inner sep = 1em, text = black] at (1.85, 0.5, 0) (conv32) {Conv2D 32};
\node[rectangle, draw, orange, fill = orange!30, inner sep = 1.4em, text = black] at (3.45, 0.5, 0) (conv64) {Conv2D 64};
\node[rectangle, draw, gray, fill = gray!30, inner sep = 1.4em, text = black] at (5.04, 0.5, 0) (maxpool) {Max Pool};
\node[rectangle, draw, gray, fill = gray!30, inner sep = 1.7em, text = black] at (6.5, 0.5, 0) (flatten) {Flatten};
\node[below = of flatten, rectangle, draw, green, fill = green!30, inner sep = 1.9 em, text = black] at (6.5, 0.5, 0) (dense128) {Dense 128};
\node[below=of dense128, rectangle, draw, green, fill = green!30, inner sep = 1.9 em, text = black] at (4.7, 0, 0) (dense5){Dense 5};

\draw[arrow] (melspec) -- (conv32);
\draw[arrow] (conv32) -- (conv64);
\draw[arrow] (conv64) -- (maxpool);
\draw[arrow] (maxpool) -- (flatten);
\draw[arrow] (flatten) -- (dense128);
\draw[arrow] (dense128) -- (dense5);
\end{tikzpicture}
	\caption{CNN model architecture for melspectrogram image classification}
	\label{fig: CNN}
\end{figure*}

Accuracy is of prime importance when comes to forensic detection like detection of synthetic speech. Deep learning models have proved to be efficient and precise in solving various complex problems. We tackled this problem by treating it as a image classification problem. The raw audio wave forms were converted to Melspectrogram images which were then passed to the CNN classifiers. An observation of importance is that the melspectrogram images of synthetic speech were different from that of the Human speech. A durable power component is missing from the synthetic speech which is a characteristic of the Human speech. Thus, using melspectrogram images is useful for the classification. The melspectrogram images have a dimension of $64\times64\times3$ which are then normalized and passed to the convolutional neural network. We have experimented with 2 different convolutional neural net architectures. One of them is a pre-trained CNN model (VGG16 \cite{vgg14}) and one is our own-built architecture. We have used weights of VGG16 \cite{vgg14} which had been trained on ImageNet except the last layer. We have added few other layers on top of VGG16 for our problem. For each model, we have used Adam optimizer with Cross Entropy loss function. For training the models, we split the dataset into 70\% training, 15\% validation, and 15\% testing.  The architecture of the model using pre-trained VGG16 model \cite{vgg14} is depicted in Fig. \ref{fig: VGG16}. The CNN architecture that we built is shown in Fig. \ref{fig: CNN}. The models were trained for multi-class classification only and were trained with early stopping to avoid any over-fitting, since the dataset we have used is not too large. The information about the hyper-parameters is presented in table \ref{table: Hyperparameters}.
\begin{table*}[ht]
\centering
\vspace{2mm}
\caption{Model hyper parameters for training}
\begin{tabular}{|c|c|c|c|}
\hline
    Model architecture &  Learning Rate & Decay & Epochs\\
\hline
VGG16 & $10^{-3}$ & $0.9$ &  155\\
Homemade CNN & $10^{-3}$ & $0.9$ &  32\\
\hline
\end{tabular}
\label{table: Hyperparameters}
\end{table*}

\section{Results and Discussion} \label{sec:results}
In  this  section,  we  have  presented  the  results  and discussion  for both  our  machine  learning  and  deep  learning based  experiments. 
\subsection{Machine Learning Based Experiments}
We tested the performance of all the aforementioned features using following  machine learning models:  Linear Discriminant, Linear SVM, Weighted KNN, Boosted Trees, Bagged Trees and RUSBoosted Trees. For binary-class classification considering the two classes as `Human Speech' and `AI Synthesized speech', the Bagged Trees classifier provided the highest accuracy of 98.9\%. Next, we experimented for a more finer multi-class classification, considering the fives classes as `Amazon Polly', `Hearling', `Human Speech', `IITM TTS', and `Voice Maker'.  Here, four classes `Amazon Polly', `Hearling', `IITM TTS', and `Voice Maker' represent the `AI Synthesized speech'.  Here again, the Bagged Trees classifier gave us the highest accuracy of 93.4\%. The confusion matrix for both the binary-class and mutli-class classification for the Bagged Trees classifier is shown in Fig. \ref{fig:binaryclasstest} and Fig. \ref{fig:multiclasstest} respectively.  Based on the  confusion matrix,  we can observe that the binary-class classification gives an higher accuracy compared to the multi-class classification. The reason for this is that there are false positives between the different classes of AI synthesized speech samples. On the contrary, in binary-class classification, all the AI speech samples are grouped into one class. This significantly  reduces the the miss classification rate. The detailed results of the accuracy of all the machine learning models that we have used is shown in Table \ref{tab:BinaryClass Experiment1} and Table \ref{tab:MultiClass Experiment1}

\begin{figure}
    \centering
    \includegraphics[width=0.5\textwidth]{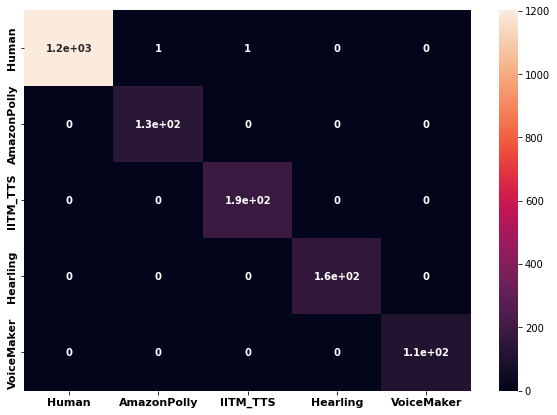}
    \caption{Confusion Matrix : VGG16 based model}
    \label{fig:vgg16confusion}
\end{figure}

\begin{figure}
    \centering
    \includegraphics[width=0.5\textwidth]{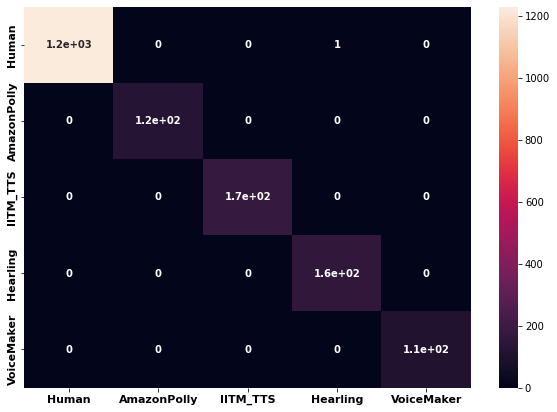}
    \caption{Confusion Matrix : Custom CNN model}
    \label{fig:customCNNconfusion}
\end{figure}

\begin{table*}[!ht]
    \centering
    \vspace{4mm}
    \caption{Deep Learning models classification results}
    \begin{tabular}{|c|c|c|c|}
    \hline
        Model Architecture &  Test Accuracy (in \%) & ROC-AUC score & F1-score (macro-average)\\
    \hline
    VGG16 & 99.83 & 0.99998 & 1.0\\
    \hline
    Homemade CNN & 99.99 & 1.0 & 1.0\\
    \hline
    \end{tabular}
    \vspace{4mm}
    \label{table: deep_l_classification_results}
\end{table*}

\subsection{Deep Learning Based Experiments}
We have employed the training using VGG16  model for 155 epochs, and obtained a validation accuracy of $99.89\%$ and testing accuracy of  $99.83\%$ respectively. The model was trained on a GPU and took about 12 minutes to train completely. The confusion matrix for the testing accuracy is depicted in Fig: \ref{fig:vgg16confusion}. 
Another deep neural net architecture that we used in our experiments is the homemade CNN. The detailed layers are shown in Fig. \ref{fig: CNN}.  We had trained the models with constant regularization using dropout layers in the architecture. It took about 5-6 minutes to train the network. The accuracy we obtained on the test data for our custom built CNN architecture was about $99.98\%$.  The confusion matrix for the classification with custom made CNN network is shown  in Fig: \ref{fig:customCNNconfusion}. Further, we computed F1 scores and ROC-AUC scores for these classification tasks to validate the obtained results. The detailed results are presented in Table \ref{table: deep_l_classification_results}.

\section{Conclusion} \label{sec:conclusion}

We found that both machine learning and deep learning based approaches perform quite well in discriminating `AI synthesized speech' from `Human speech'.  The accuracy achieved using CNN architecture was about $99.99\%$ that was slightly higher than that achieved by the Bagged Trees classifier ($98.9\%$). The time taken to achieve the aforementioned accuracy was higher for CNN architecture compared to time for the Bagged Trees classifier. In near future, we want to put efforts towards reducing this miss classifications and also, experiment with other case scenarios like identifying synthesized male versus female speech, classifying based on age groups, and various other aspects.  

% we saw that our models of deep learning outperformed our machine learning model based approach and gave a better accuracy. But given the time of training which was less for our machine learning approach, we can see the accuracy achieved by our handcrafted features in machine learning for binary class classification is giving at par results. We believe that given the case scenarios, both our techniques can act as a good agent to detect the AI synthesized speech depending upon the application.

\section{References}
\renewcommand{\section}[2]{}%
\bibliographystyle{plain}
\bibliography{references.bib}
\vspace{12pt}
\color{red}

\end{document}